\documentclass[preprint2]{aastex631}
\usepackage{graphicx,times}             %for PS/EPS graphics inclusion, new
\usepackage{color}

\usepackage{ulem}

\newcommand{\shao}{Shanghai Astronomical Observatory, Chinese Academy of
		Sciences, Shanghai 200030, China}
\newcommand{\nav}{Shanghai Key Laboratory of Space Navigation and Positioning 	
	Techniques, Shanghai 200030, China}	
\newcommand{\dc}{National Basic Science Data Center, Beijing 100190,
China}
\newcommand{\apshao}{Key Laboratory for Research in Galaxies and Cosmology, Shanghai Astronomical Observatory,\\ Chinese Academy of Sciences, 80 Nandan Rd., Shanghai, 200030, China}

\definecolor{blue}{rgb}{0.0,0.0,1.0}
\newcommand{\revI}[1]{#1}
\newcommand{\revII}[1]{#1}
\newcommand{\uvom}{Omni\textsl{UV}}

\submitjournal{AJ}

\begin{document}
	
\title{\uvom: A Multi-Purpose Simulation Toolkit for VLBI
Observation}

\author{Lei Liu}
\affiliation{\shao}\affiliation{\dc}\affiliation{\nav}

\author{Weimin Zheng}
\affiliation{\shao}\affiliation{\dc}\affiliation{\nav}

\author{Jian Fu}
\affiliation{\apshao}

\author{Zhijun Xu}
\affiliation{\shao}

\correspondingauthor{Lei Liu}
\email{liulei@shao.ac.cn}

\begin{abstract}We present \uvom, a multi-purpose simulation toolkit
for space and ground
VLBI observations. It supports various kinds of VLBI stations,
including Earth (ground) fixed, Earth orbit, Lunar fixed,
Lunar orbit, Moon-Earth and Earth-Sun Lagrange 1 and 2 points, etc. The main
functionalities of this toolkit are: (1) Trajectory calculation; (2)
Baseline $uv$ calculation, by taking the availability of each
station into account; (3) Visibility simulation for the given $uv$
distribution, source structure and system noise; (4) Image and beam
 reconstruction. Two scenarios, namely \revI{the} space VLBI network and \revI{the} wide
 field
array, are presented as \revI{the} application of the toolkit 
at completely different scales.
\uvom~is the acronym of ``Omnipotent \textsl{UV}''. We hope it could
work as a general framework, in which various kinds of stations could be
easily incorporated and the functionalities could be further extended.
The toolkit has been made publicly available.
\end{abstract}


\section{Introduction}
Data simulations are important for the schedule and evaluation of 
radio interferometric observations
\citep{pyuvsim}.
RIME (Radio Interferometry Measurement Equation), formulated by
\citet{Hamaker1996}, provides the basic mathematical framework that
links the observed source with the finally
recorded signal. Various propagation effects, e.g., phase delays,
parallactic angle rotations, receiver gains, beam patterns, could be
incorporated into the framework in an elegant and elaborate way
\citep{Smirnov2011}.
Since its emergence, most of simulation and calibration tools
are developed under this framework:
%Besides that, each tool has its own focus on the series of
% terms (Jones matrix), so as to fulfill the requirements of
% the corresponding instruments.
\texttt{OSKAR} \citep{oskar} is an interferometer and
 beam forming simulator package
dedicated to \revI{the} simulation of SKA \citep[Square Kilometer Array, ][]{ska}.
It implements
a hierarchical structure, in which both of the antenna field
pattern within a station and the station beam are carefully
modeled. The use of GPU (Graphics Processing Unit) makes it possible 
to support the large
number of pixels and stations which are required by SKA simulation.
\texttt{pyuvsim} \citep{pyuvsim} is another visibility simulation
 package with an emphasis on
accuracy and design clarity over efficiency and speed, so as to
achieve the necessary high level precision for neutral hydrogen studies.
It runs on CPU clusters and uses MPI (Message Passing Interface) for
parallelization.
\revII{Besides \texttt{pyuvsim}, the ``Radio Astronomy Software Group''\footnote{https://github.com/RadioAstronomySoftwareGroup} also
maintains \texttt{pyuvdata} and
\texttt{pyradiosky}, 
which are indispensable for radio interferometry
simulations.} \texttt{CASA} \citep[the Common Astronomy Software Applications package, ][]{casa}
is the primary data processing software for ALMA (the Atacama Large Millimeter/sub-millimeter
 Array) and VLA (Very Large Array). It also
provides tools for visibility simulation based on RIME. \texttt{CASA} is build
upon a set of C++ libraries (``CasaCore'') and use Python for its
interface, which guarantee both efficiency and \revI{user-friendliness}.
%SYMBA is an end-to-end VLBI synthetic data generation pipeline
%dedicated to the simulation of EHT observations. It uses
%\texttt{MeqSilhouette} for raw data generation, and then uses
%rPICARD for VLBI data calibration.
Another tool \revII{worth mentioning} is \texttt{MeqTrees} \citep{meqtree}.
It is designed to be able ``to implement an arbitrary
measurement equation and to solve for arbitrary sets of its parameters''.
To achieve that, a Python-based Tree Definition Language (TDL) is
designed and realized. Based on \texttt{MeqTrees}, a new package,
\texttt{MeqSilhouette} \citep{meqsil}
\revI{was} developed. It is specifically designed for the accurate simulation of
\revI{the} EHT \citep[Event Horizon Telescope, ][]{EHT} observation. Based on that,
a series of signal corruptions that are
important in millimeter wavelength, including troposphere, ISM (Interstellar
medium) scattering,
and time-variable antenna pointing errors, are taken into account.

All tools mentioned above are intended for ground based telescopes.
In principle, if baseline $uv$s are available, \revI{those tools could} be
used for \revI{space VLBI simulations} as well. However,
space VLBI involves calculation of many kinds of
trajectories. E.g., Earth orbit, Lunar orbit, Lagrange points, etc.
Although all the necessary methods and equations for those calculations
are \revI{described} in the standard textbooks, and all the required ephemeris
are publicly available, the actual implementation is still not a trivial
task.
Moreover, visibility simulation requires not only trajectories, but also
the availability of each telescope.
For ground based telescopes, availability is mainly determined by the minimum
elevation angle. For space telescopes, \revI{it also involves the the minimum 
separation angle between the source and the celestial object being considered.}
Simulations of space VLBI observations are still in its preliminary stage. 
\citet{Andrianov2021} \revI{demonstrated} the improvement of image resolution for the
joint observations with Millimetron \citep{MM} and EHT.
\citet{Palumbo2019} \revI{investigated} the possibility of detection for the 
black hole rapid time variability
by expanding the array to include telescopes in LEO (Low Earth Orbit).
\revI{Also note that the ``Fakerat'' software package\footnote{http://www.asc.rssi.ru/radioastron/software/soft.html} provides support for 
observation planning, $uv$ calculation and beam/image reconstruction of the 
RadioAstron mission. However, it is dedicated to that mission only and 
requires the satellite orbit as input.} According to our investigation, 
\revI{at present, no general purpose simulation tools are publicly available 
for space VLBI observations.}

Several space VLBI projects are planned or already
under development. China is planning its first Moon-Earth space VLBI
experiment in the Chang'E 7 mission. This is based on the 4.2 meter relaying
antenna and will perform VLBI observations in X-band.
The lunar surface radio telescope is also at discussion.
At present, SHAO (Shanghai Astronomical Observatory, Chinese Academy of Sciences)
is proposing the Space Low Frequency Radio Observatory, of which
two satellites each equipped with a 30 meter radio telescope will be sent to the
Earth elliptical orbit (orbit height 2,000~km $\times$ 90,000~km). For the next
stage of the Event Horizon Telescope, a very
natural extension is to include space radio
telescopes, so as to achieve even higher angular resolution.
All of those projects require appropriate tools for the simulation of the
corresponding VLBI observations.

% Influence of w term, gridding artifacts

Keeping this in mind, we develop \uvom \citep{omniuv}, so as to fulfill the
requirement of various kinds of VLBI observations. At present,
\uvom~provides the following functionalities:
\begin{itemize}
\item Trajectory calculation for various kinds of stations;
\item $uvw$ and telescope/baseline availability calculation;
\item Visibility simulation based on the given source structure and 
$uvw$, by taking the influence of system noise into account;
\item Image and beam pattern reconstruction, so as to provide
appropriate tools for the
evaluation of observation quality with given configuration.
\end{itemize}
One thing we want to point out is, for visibility simulation and
radio imaging, \uvom~implements both FFT (Fast Fourier Transform)
and \revI{DFT (Discreate Fourier Transform)} methods.
The latter one \revI{makes it possible to support the $w$ term}. For 
wide field imaging, the
variation of phase caused by $w$ term must be taken into account, so
as to avoid the distortion of the resulting wide field image. Moreover,
FFT method requires gridding,
which might introduce errors that are significant for the data processing
of 21~cm experiments \citep{Trott2012}. In this work, we will demonstrate
the simulation results for configurations of both space VLBI network
and wide field array.

This paper is organized as follows: Sec.~\ref{sec:implement} describes the
detailed implementation of the toolkit, including trajectory calculation,
$uv$ calculation, visibility simulation and image reconstruction;
Sec.~\ref{sec:demo} demonstrates the application of \uvom~in two typical
observation scenarios;
Sec.~\ref{sec:future} discusses possible future work; Sec.~\ref{sec:sum}
 presents the summary.

\section{Implementation}\label{sec:implement}
\begin{figure*}
\plotone{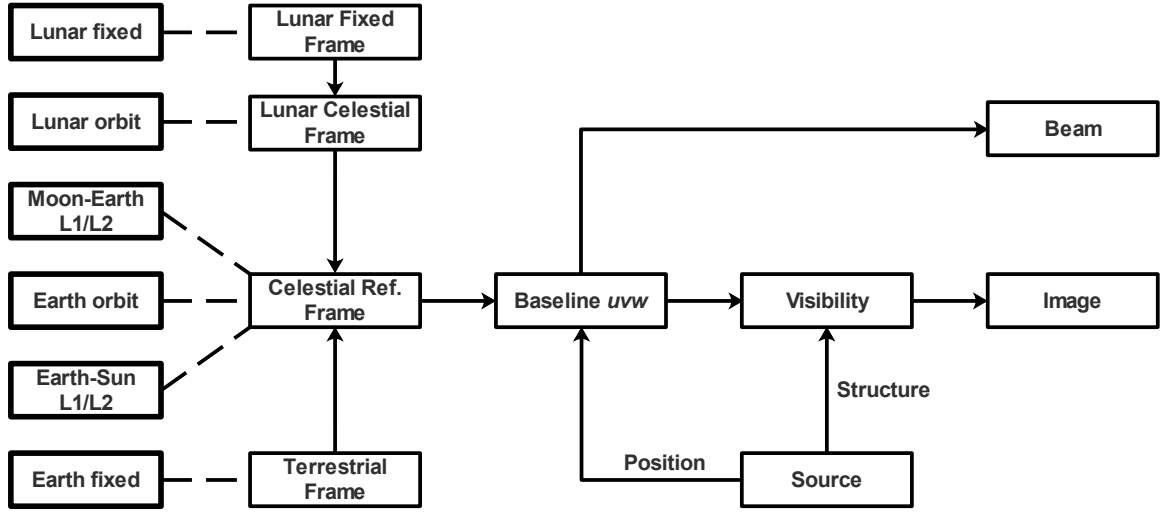}
\caption{Data flow of the \uvom~toolkit.\label{fig:dataflow}}
\end{figure*}

Fig.~\ref{fig:dataflow} demonstrates the data flow of the \uvom~toolkit.
At present,  8 types of stations are supported.\footnote{For \revI{the} Earth-Sun and \revI{the} Moon-Earth system,
stations in \revI{the} Lagrange 1 and 2 points are supported.} For the given trajectory,
baseline $uvw$ is calculated. In this process the availability of each telescope is taken
into account. Visibilities are calculated for each $uvw$ sampling point.
The radio image is constructed accordingly.

\subsection{Trajectory calculation}
Trajectory calculation involves a series of coordinate transformations.
Positions of all kinds of stations are
unified in the same frame: the Celestial Reference Frame (CRF).
The availability of each station is determined by the angular distances between
the source and the celestial objects (Earth, Moon, Sun, etc.). For \revI{the} Earth/Moon
fixed stations, elevation angles are considered as well. In \uvom, the
minimum distance for each object and the minimum elevation angle can be
configured explicitly.

\subsubsection{Moon stations}
\uvom~utilizes the JPL Planetary Ephemeris (version DE-421) Lunar PCKs, which
provides the orientation of \revI{the} Lunar Principal Axis (PA). The Euler angles at \revI{the} given
solar system barycenter Julian date (TDB) are loaded, so as to build the rotation
matrices for the conversion from LFF (Lunar Fixed Frame) to LCF (Lunar Celestial
Frame). The relative position of \revI{the} Moon
to \revI{the} Earth is retrieved from the JPL Planetary Ephemeris (version DE-421) SPK.
In this way the coordinates of \revI{the} Moon fixed and \revI{the} orbit stations in CRF are calculated.

\subsubsection{Earth stations}

\revI{The trajectories} of \revI{the} Earth orbit stations are described with 6 orbital elements 
and are calculated in CRF. \revI{The trajectory} calculation of \revI{the} Earth fixed stations 
involves translation from TF (Terrestrial Frame) to CRF,
which requires \revI{the} matrices of polar motion (wobble, $\mathbf{W}$), Earth rotation
($\mathbf{R}$), precession and nutation ($\mathbf{PN}$). These calculations 
require \revI{the} EOP (Earth Orientation Parameter), \revI{which are updated on a daily 
basis. Since the retrieve of \revI{the} EOP might not be a trivial task for some users, 
the input of \revI{the} EOP is optional. It is the decision of the user to provide
EOPs at specific dates to achieve the necessary precision of \revI{the} trajectory calculation.}

\subsubsection{Stations at Lagrange points}

The location of the L1 point is the solution to the following equation:
\begin{equation}\label{eq:L1}
\frac{M_1}{(R-r)^2}=\frac{M_2}{r^2}+\left(\frac{M_1}{M_1+M_2}R-r\right)
                    \frac{M_1+M_2}{R^3},
\end{equation}
where $r$ is the distance of \revI{the} L1 point to the smaller object, $R$ is the
distance between the two objects, $M_1$ and $M_2$ are the masses of the
large and \revI{the} small object, respectively. Given $R$, $M_1$ and $M_2$, $r$
could be solved numerically.

The location of \revI{ the L2 point} is the solution to the following equation:
\begin{equation}\label{eq:L2}
\frac{M_1}{(R+r)^2}+\frac{M_2}{r^2}=\left(\frac{M_1}{M_1+M_2}R+r\right)
                    \frac{M_1+M_2}{R^3},
\end{equation}
with parameters defined as that in Eq.~\ref{eq:L1}.

The relative positions of \revI{the} Moon and \revI{the} Sun to \revI{the} Earth are retrieved from
the JPL Planetary Ephemeris (version DE-421). Noting that the L1/L2
points lie on the line defined by the two celestial objects, their
positions could be derived with the corresponding $r$, as listed
in Tab.~\ref{tab:rval}.

\begin{table}
\centering
\caption{The L1/L2 $r$ values of the Moon-Earth and \revI{the} Earth-Sun system. $r$
is given in unit of $R$ (distance between the two celestial objects).
\label{tab:rval}}
\begin{tabular}{ccccc}
\hline\hline
&\multicolumn{2}{c}{Moon-Earth} & \multicolumn{2}{c}{Earth-Sun} \\
%\cline{2-5}
\hline
&L1 & L2 & L1 & L2 \\
\hline
$r$ &0.15091 & 0.16780 & 0.00997 & 0.01004 \\
\hline
\end{tabular}
\end{table}

\subsection{$uvw$ calculation}
Once the trajectory of each station is obtained, the corresponding $u, v, w$ 
calculations are straightforward. 
\revI{The definition of the $uvw$ system in \uvom~follows that in the standard 
textbook \citep{Thompson2001}:
given the North Pole direction $\vec{n} = (0, 0, 1)$ and the source direction 
$\vec{s}_0$,  $\vec{e}_w$ takes the same direction of $\vec{s}_0$:}
$$\vec{e}_w = \vec{s}_0,$$
\revI{$\vec{e}_u$ is the direction perpendicular to the plane defined by 
$\vec{n}$ and $\vec{e}_w$:}
$$\vec{e}_u = \vec{n} \times \vec{e}_w,$$
\revI{$\vec{e}_v$ is defined accordingly:}
$$\vec{e}_v = \vec{e}_w \times \vec{e}_u.$$
\revI{$u, v, w$ of a given baseline are the projections of the baseline 
vector $\vec{b}$ in the $uvw$ system:}
$$(u, v, w) = \vec{b} \cdot (\vec{e}_u, \vec{e}_v, \vec{e}_w),$$
One thing that must be taken into account is the telescope availability. 
For Earth and Moon fixed stations, this involves the calculation of the 
minimum elevation angle. For Space stations,
this is determined by the minimum separation angle between the observed
source and the celestial object. \revI{All parameters mentioned above, together
with the celestial objects being considered,}
could be set in \uvom. The toolkit will take care
of all the necessary calculations.

\subsection{Visibility simulation}\label{sec:vis}
The mathematical basis that connects the observed visibilities and
the intrinsic source brightness is RIME, of which various propagation
effects are described
by the corresponding Jones matrices \citep{Smirnov2011}.
Among them, the $K$ Jones that
describes the phase delay is at the heart of interferometry \citep{meqtree}.
In this framework, by assuming the observed source is composed of a
series of point sources, the visibility of \revI{the} given baseline $k$ could be
expressed as:
\begin{equation}\label{eq:vis}
V_k = \sum_{i} S_i e^{-j2\pi(u_k l_i + v_k m_i + w_k (n_i-1))},
\end{equation}
where $u_k, v_k, w_k$ are \revI{the} projections of the baseline in the $uvw$
coordinate system, $S_i$ are the flux intensity of the $i$~th point,
$l_i, m_i, n_i=\sqrt{1-l_i^2-m_i^2}$ are the corresponding
direction cosines.

\uvom~provides two methods for visibility calculation at each
$uvw$ sampling point. The first one is the FFT based
method. In this method, the flux intensities of point sources are
first mapped to a 2-D array of equally spaced grids in the image
plane. The grid size is
selected according to the angular resolution of the VLBI network.
After that this array is transformed to the $uv$ plane with FFT.
Visibilities are obtained at each $uv$ grid. Finally baseline
visibilities are reconstructed based on their $uv$ position. 
\revI{In the current treatment of \uvom, the flux of each point source
is assigned to the nearest grid (pixel) in the image plane, and each 
visibility sampling point takes the value of 
the nearest grid in the $uv$ plane.} The main purpose of gridding is 
to use FFT, which greatly reduces \revI{the}
computational complexity. This guarantees that \revI{the} visibility
calculation could be \revI{performed} with limited hardware
resources, which \revI{were} crucial in the past forty years. 
However, \revI{the} gridding
process involves flux intensity assignment to the grids with
certain assignment method, which introduces the assignment window
function. 
\revI{This} window function convolves with the actual
visibility in the $uv$ plane \revI{after FFT}, which can be regarded as 
extra noises (artifacts). For certain kind of experiment that requires 
high precision, this artifact is non-negligible \citep{Trott2012}. 

Another method is \revI{DFT (Discrete Fourier transform)}, which is based 
on Eq.~\ref{eq:vis}.
Unlike the FFT method, the influence of $w$ term is incorporated to
the visibility calculation, \revI{which is important for the simulation of 
wide field array}. Besides that, since
it does not require gridding, the artifact of windowing function could
be completely avoided. Both of above guarantee a more accurate
visibility calculation. \revI{However, we have to point out that the DFT method 
comes at the price of a much larger computational cost, which makes it 
almost impossible for actual applications without the aid of hardware 
accelerators.}

In \uvom, for visibility calculation, the effects of thermal noise 
and antenna
gain are taken into account. The implementation follows that in
\citet{Chael2018}. The full complex visibility is expressed as:
\begin{equation}
V_{ij} = G_i~G_j~e^{-j(\phi_i-\phi_j)}(V_{0, ij} + \epsilon_{ij}),
\end{equation}
where $V_{0, ij}$ is the simulated visibility according to the
theoretical expression in Eq.~\ref{eq:vis}.
$G_i = \sqrt{1+Y_i(t)}$ is the antenna gain of station $i$,
$Y_i$ is the Gaussian random variable with zero mean, and is drawn
each time $uvw$ is sampled. Following \citet{Chael2018}, the
standard deviation for the Gaussian distribution of $Y_i$ is named
as gain error, and is regarded as a free parameter.
The additional phase at each station depends on the atmospheric
coherence time. For high frequency observation, it can be regarded
as a uniform distribution in the range $-\pi<\phi<\pi$ and \revI{is} sampled
for each visibility. For low frequency and space observations, the
phase is constant. The thermal noise $\epsilon_{ij}$ is calculated
for each visibility measurement and is described as a complex random
Gaussian with zero mean. Its standard deviation is determined with:
\begin{equation}
\sigma_{ij} = \frac{1}{\eta}\sqrt{\frac{\mathrm{SEFD}_i\times\mathrm{SEFD}_j}{2~B~T}},
\end{equation}
where $\eta$ is the digital quantization loss, which takes the
value 0.88 for 2 bits quantization \citep{Thompson2001}.
``SEFD'' is the system
equivalent flux density, which is a measurement of the antenna
performance. $B$ and $T$ are the bandwidth and the integration
time of the visibility measurement.

\subsection{Image reconstruction}
Radio imaging is the reverse process of visibility calculation.
Given the visibility measurements of $N$ baselines, the brightness
of point source $S_i$ is:
\begin{equation}\label{eq:img}
S_i = \sum_{k}V_k~e^{j2\pi(u_k l_i+v_k m_i + w_k(n_i-1))}/N,
\end{equation}
with parameters defined in Eq.~\ref{eq:vis}

\uvom~provides both FFT and DFT methods for radio imaging, which is 
the same as that for visibility calculations.
Moreover, \uvom~is able to generate the beam pattern
for the corresponding $uv$ distribution. One can then use the method
described in \citet{Liu2021} to evaluate the observation
quality.

\begin{figure}
\plotone{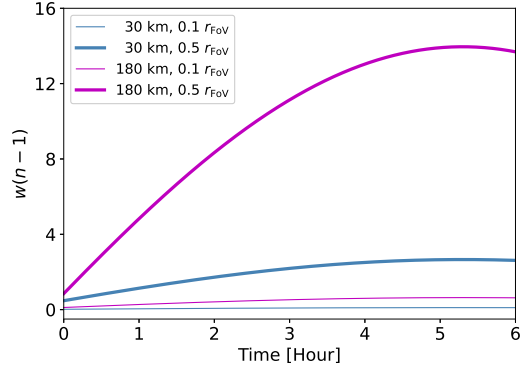}
\caption{The evolution of $w(n-1)$ term at different \revI{parameter combinations}.
    Steel blue and magenta
     correspond to the baseline lengths of 30~km and 180~km, respectively.
     Thin and thick lines correspond
     to the different offsets (as a fraction of FoV radius) to the phase center.
	\label{fig:n1w}}
\end{figure}

One thing we must pay special attention is the $w$ term.
For space VLBI observations that focus on high resolution and are always
small field, this term is negligible. However, for
wide field imaging, the influence of
this term is large and must be taken into account. Fig.~\ref{fig:n1w} demonstrates
the contribution of $w$ term to the propagation phase at different
distances to the phase center and baseline lengths. For long baselines and
large offset (to the
phase center), the influence of this term is clearly exhibited.
The neglection of $w$ term in FFT method introduces
large phase error, which leads to the distortion of \revI{the} image and 
\revI{the} decrease of \revI{the} dynamic range.

According to Eq.~\ref{eq:img}, for DFT, it is quite natural
to incorporate the contribution of $w$ term into the reconstruction of each
pixel and therefore overcomes the difficulties mentioned above.
However, compared with FFT method, this method is more computational expensive,
which makes it almost unaffordable with a single CPU. Nowadays, with the fast
development of modern hardware accelerators, the application of this
method in real simulations becomes possible. Moreover, with the aid of
modern tensor libraries and computing frameworks, the implementation of the
DFT method is rather easy. In the radio imaging part of
the current version of \uvom, Eq.~\ref{eq:img} has been implemented using both
NumPy and CuPy. The demonstration of this method is presented in
Sec.~\ref{sec:widefield}.

\section{Demonstration}\label{sec:demo}
The purpose of this section is to demonstrate the capability of \uvom~under
different scenarios. Two simulations of quite
different scales, namely space VLBI network and wide field array, are presented.

\begin{table*}
\centering
\caption{\revI{Configurations of station} for the proposed space VLBI observations
in Sec.~\ref{sec:svlbi}. \revI{$a$ (semi-major axis), $e$ (eccentricity), 
$i$ (inclination), $\Omega$ (longitude of the ascending node), $\varpi$ 
(argument of periapsis) and $M_0$ (true anomaly) are the 6 orbital elements 
for each space telescope.}}
\label{tab:station_config}
\begin{tabular}{lll}
\hline\hline
ID & Station & Description \\
\hline
1 & Earth orbit & SEFD~=~~225~Jy, D~=~~30~m, $a$~=~6.14$\times10^4$~km, $e$~=~0.73, $i$~=~~30$^\circ$, $\Omega$~=~0, $\varpi$~=~0, $M_0$~=~0\\
2 & Earth orbit & SEFD~=~~225~Jy, D~=~~30~m, $a$~=~6.14$\times10^4$~km, $e$~=~0.73, $i$~=~-30$^\circ$, $\Omega$~=~0, $\varpi$~=~0, $M_0$~=~180$^\circ$\\
3 & Earth fixed & SEFD~=~~~~48~Jy, D~=~~65~m, TMRT (Tianma Radio Telescope)  \\
4 & Earth fixed & SEFD~=~~~~20~Jy, D~=100~m, Effelsberg Radio Telescope \\
%\hline
5 & Lunar orbit & SEFD~=~~507~Jy, D~=~~20~m, $a$~=~3$\mathrm{R}_\mathrm{Moon}$, $e$~=~0, $i$~=~0$^\circ$, $\Omega$~=~0, $\varpi$~=~0, $M_0$~=~0\\
6 & Lunar fixed & SEFD~=2028~Jy, D~=~~10~m, far side of the Moon\\
7 & Moon-Earth L2 & SEFD~=~~507~Jy, D~=~~20~m\\
\hline
\end{tabular}
\end{table*}

\begin{figure*}
\plotone{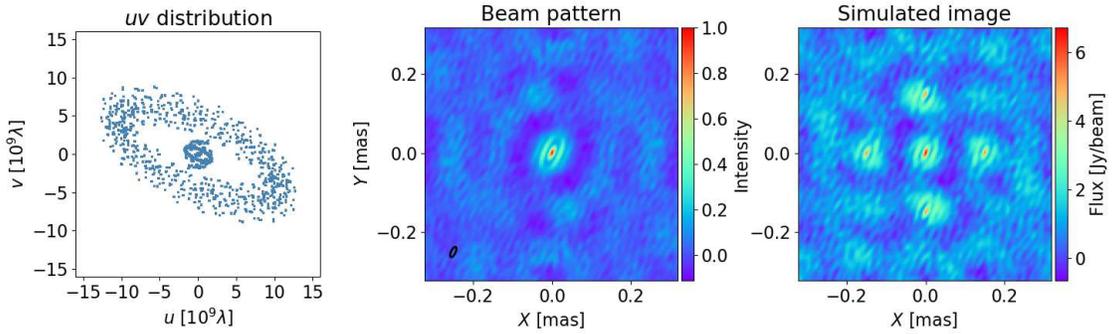}
\caption{Simulation result for the Moon-Earth network proposed in 
Sec.~\ref{sec:svlbi}. Left: $uv$ distribution. Middle: beam pattern.
Right: reconstructed image. \revI{The beam shape derived with TPJ's algorithm is 
presented as a black circle in the lower left corner of the middle panel. 
Beam size: 
major 0.028~mas, minor 0.011~mas. The input point source distribution 
is demonstrated in Fig.~\ref{fig:sim_input}.}
\label{fig:sim_me}}
\end{figure*}

\begin{figure*}
\plotone{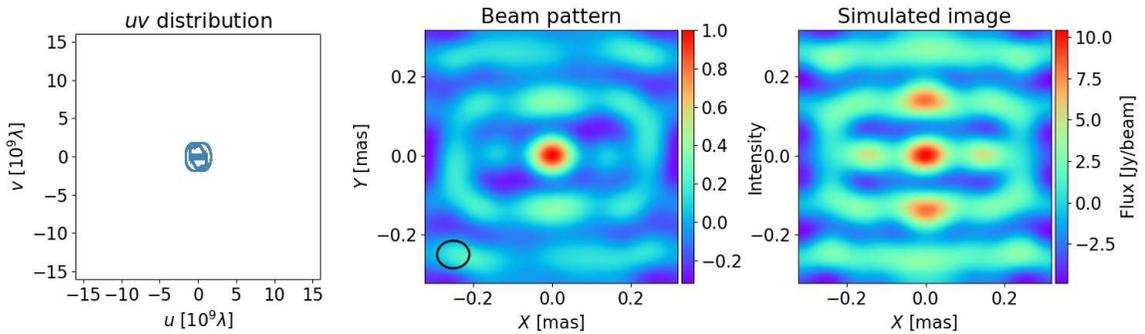}
\caption{Simulation result for the Earth only network proposed in
Sec.~\ref{sec:svlbi}. The meaning of each panel is the same as that 
in Fig.~\ref{fig:sim_me}. Beam size: major 0.082~mas, minor 0.070~mas. 
\label{fig:sim_e}}
\end{figure*}

\begin{figure}
\plotone{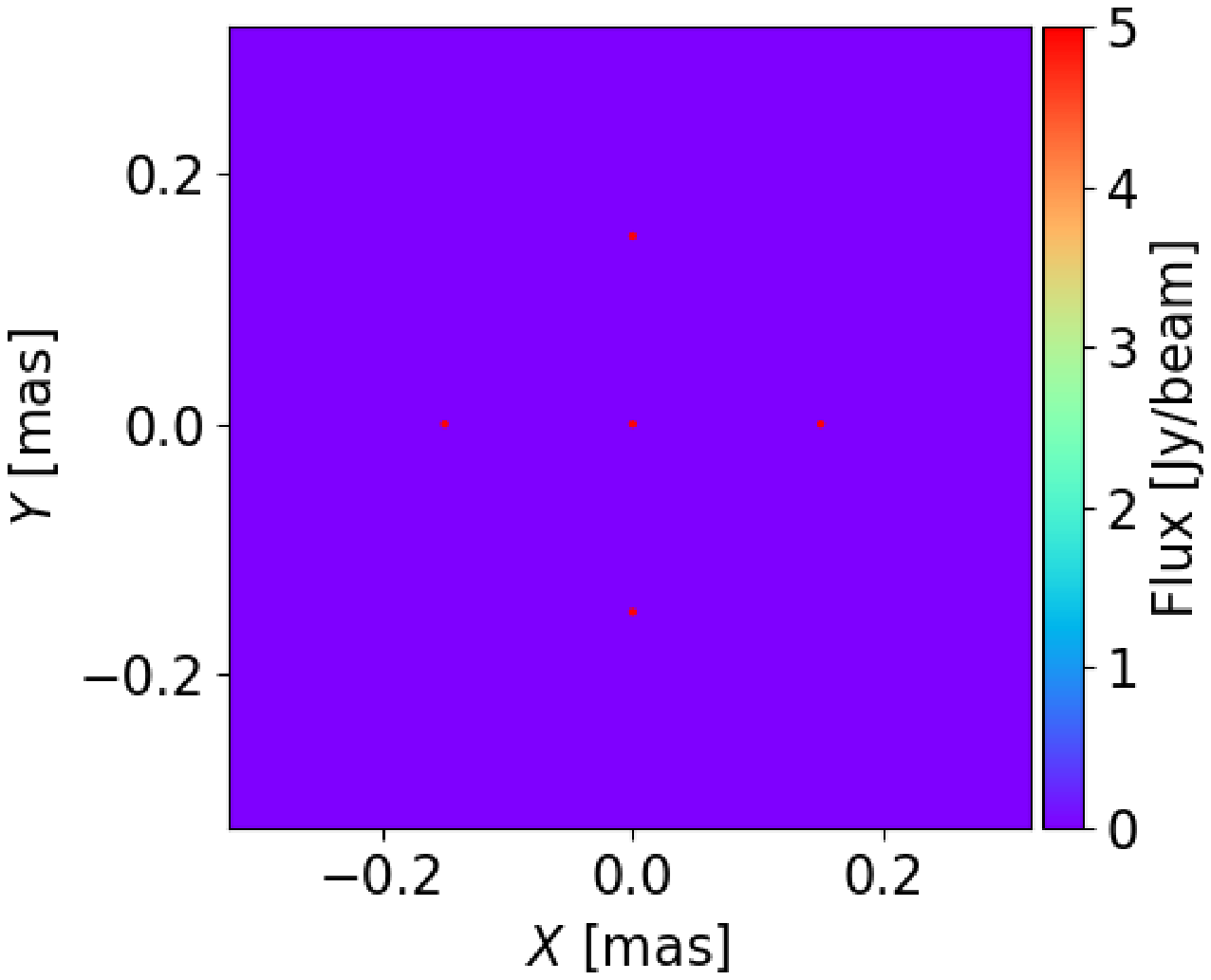}
\caption{\revI{The input structure for space VLBI simulations in 
Sec.~\ref{sec:svlbi}. 5 point sources with a separation of 
0.15~mas are placed in the center of the field of view (FoV). The 
flux of each point source is set to 5~Jy. Note that the dots 
in the figure do not indicate the size of ideal ``point''
sources.}\label{fig:sim_input}}
\end{figure}

\subsection{Space VLBI Network}\label{sec:svlbi}
This simulation demonstrates the improvement of \revI{the} angular resolution
and \revI{the} image quality with and without \revI{the} Moon related baselines.
The image reconstruction results of \revI{the} Earth only and \revI{the} Moon-Earth
VLBI network are presented and compared.

%\subsubsection{Simulation setup}
The Earth only space VLBI network consists of two ground based
and two space stations. For the Moon-Earth network,
as a demonstration, three Moon stations, each deployed in
the Moon orbit, the farside of the Moon surface and
the Moon-Earth L2 point, are included. Tab.~\ref{tab:station_config}
presents the detailed configurations of the proposed stations. 
The main parameters of the observations are: frequency: X band (3.6~cm), 
phase center: R.A. 180$^\circ$, Dec. 30$^\circ$, bandwidth: 32~MHz,
integration time: 2~s, gain error: 0.1.

The source structure is simulated
with 5 point sources. The distance between the sources
is set such that the source structure could be resolved
by both of the Earth only and the Moon-Earth
network. As a raw estimation, the baseline length of \revI{the} Earth
only network is in the level of $10^5$~km. In \revI{the} X band, the
corresponding angular resolution is about 0.07~mas. As a result,
the distance between the sources is set to 0.15~mas.
% Earth only: 6.46E4 km, 0.114 mas @ 8.4 GHz
% Moon-Earth: 45.31E4 km, 0.016 mas @ 8.4 GHz

%\subsubsection{Schedule}
One thing that we must pay special attention is
the schedule of space VLBI observations. For \revI{the} Earth only network, since
there are two ground based stations, to
achieve the best $uv$ coverage, the total duration is set to
one day, which is the same as the ground only VLBI session.
For the Moon-Earth network, we have
to realize that the situation is quite different.
For one thing, the revolution period of the Moon is one month
instead of one day. Moreover, Moon-Earth baselines are much
longer than Earth only baselines.
Keeping these in mind, we propose the following observation strategy:
\begin{itemize}
\item Observations are organized in scans. Each scan lasts
for tens of minutes.
\item Several scans are arranged within one day.
\item Several days are selected for \revI{the} observation within one month.
\end{itemize}
This strategy guarantees that the high resolution of Moon-Earth
baselines could be fully utilized with relatively good $uv$ coverage.
For the observation presented in this section, the configuration
is 15 minutes per scan, 2 scans per day with an interval of 12 hours,
28 continuous days.

%\subsubsection{Simulation result}
\revI{The input structure is demonstrated in Fig.~\ref{fig:sim_input}.}
The simulation results with and without Moon related baselines
are presented in Fig.~\ref{fig:sim_me} and Fig.~\ref{fig:sim_e},
respectively. For convenience, the beam size calculated with
the TPJ's algorithm\footnote{\revI{Please refer to DIFMAP 
\citep{DIFMAP} for more details.}} is presented for \revI{the} beam
pattern. Although 5 points could be resolved in both observations,
the Moon-Earth network \revI{yields} much higher \revI{angular} resolution, and
therefore \revI{resolves more details}. Moreover, the relatively
better $uv$ coverage of the Moon-Earth network exhibits smaller
side lobe. As a result, the corresponding image is reconstructed
with higher quality.

\subsection{Wide field array}\label{sec:widefield}
%\subsubsection{Simulation setup}
The simulation of wide field array is based on the planned SKA1-Mid array. 
For visibility
simulation and image construction, DFT method is used,
so as to include the contribution of $w$ term and to avoid the artifact
of gridding. As pointed out in Sec.~\ref{sec:vis}, DFT is computationally
expensive. As a result, the number of antennas that take part in observation
is constrained:
% 190 15 meters, 153 within 15~km, 20 selected.
instead of using all antennas in the planned array,
20 out of 190 (15~m diameter) antennas are selected randomly within a 
radius of 15~km
in the central area of the array. The actual size of the array leads to
an estimated angular resolution of 13.4 arcsec at 350~MHz. The corresponding 
pixel size is set to 7.5 arcsec. As a modeling
of the neutral hydrogen foreground, 20 point sources with a flux of 1~Jy
are placed randomly in the FoV. The main parameters of the
observations are: frequency: 350~MHz, phase center: R.A. 180$^\circ$, Dec.
-60$^\circ$, bandwidth: 32~MHz, integration time: 2~s, gain error: 0.3, 
duration: \revI{6 hours}. 
%Considering the minimum elevation angle of the antenna
%is set to 15 degree, the total observation time is around 15 hours.

% 280 s, 190 baselines, 2048^2 pixels, 174901 vis

\begin{figure*}
\plotone{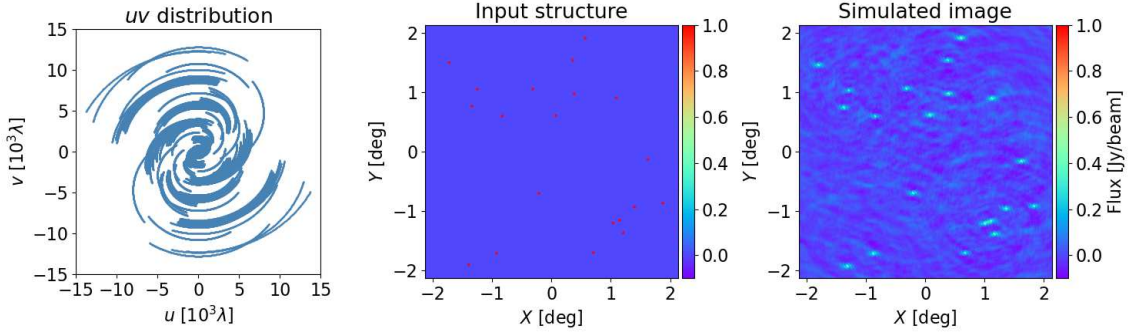}
\caption{Simulation result for the SKA1-Mid array proposed in 
Sec.~\ref{sec:widefield}. \revI{Left: $uv$ distribution. Middle: input 
point source distribution. Right: reconstructed image. 
Beam size: major 27.9~arcsec, minor 19.6~arcsec. As a modeling of 
the neutral hydrogen foreground, 20 point sources with a flux of 
1~Jy are placed randomly in the FoV. Note that the dots 
in the middle panel does not indicate the actual size of 
ideal ``point'' sources.}\label{fig:sim_ska}}
\end{figure*}

Fig.~\ref{fig:sim_ska} presents the simulation result. 20 antennas yield
190 baselines and therefore relatively good $uv$ coverage. We may expect the image
quality will be further improved by taking all antennas of SKA1-Mid into account.

One may notice that both scenarios presented in this section are 
simulations for point sources. Actually we have carried out simulations 
for diffuse structures as well.
The result proves that the source structure could be reconstructed according to
the given beam. We hope \uvom~could be used to simulate the 21~cm signal by 
interferometric observations from HI gas in galaxies in the nearby universe, 
which is one of the main science goals for SKA1-Mid \citep{2015aska}.
We also realize that since structures smaller than the beam
size could not be resolved, the sampling resolution could be set
accordingly, so as to reduce the total number of sampling points in the
image plane and therefore the amount of computation.

\section{Future work}\label{sec:future}
As a novel VLBI simulation toolkit, \uvom~is still in its
preliminary stage. Several new features are already planned
and will be implemented in the near future.

\subsection{DRO}
With the development of modern space technology, a special
kind of orbit, namely DRO (Distant Retrograde Orbit), has obtained
more and more attention. DRO is a family of periodic orbits
in the circular restricted three-body problem (CR3BP). In the
rotating reference frame, it looks like a retrograde orbit
around the second body. DRO is proved to be long-term stable,
and is therefore regarded as the ideal place for various kinds
of space missions. It is necessary to investigate
the possibility of the deployment of space VLBI satellite in this
orbit. At present, DRO is not yet supported by \uvom. The main
reason is that the object in this orbit is influenced by both the
primary and the
secondary body, which makes it impossible to derive any analytical
solution. As a result, the trajectory of DRO can only be calculated
numerically. \citet{Zimovan2017} provides an excellent introduction
to DRO and has listed all the possible initial conditions (position
and velocity) for the DRO family in the Moon-Earth system.
As a next step, a high accuracy numerical integrator
will be incorporated to \uvom, so as to provide \revI{the} support for
the trajectory calculation of DRO and the related visibility and
image simulations.

\subsection{Performance improvement}
The computational complexity of FFT method is low. For most of \revI{the} space VLBI
simulations using this method, the time consumption is already acceptable
with only one CPU. However, wide field imaging requires DFT
 to include the contribution of $w$ term, of which the computational
requirement is much higher. The SKA1-Mid simulation presented in Sec.~\ref{sec:widefield}
takes 5 minutes on 3 GPUs. Note that it includes only 20 
antennas within a radius
of 15~km in the array center. The actual size of the array is 180~km,
which suggests 1 order of magnitude higher angular resolution. The
corresponding 2 orders of magnitude increase in the computation requirement is
only possible with modern GPU clusters. Therefore, DFT
part is planned to be fully parallelized with MPI and accelerated with
multiple types of hardware backends.

\subsection{Data corruption}
A realistic simulation must take various data corruption effects
into account. In the current version of \uvom, \revI{the} antenna gain and
\revI{the} system noise have been incorporated by following the
scheme adopted by \citet{Chael2018}. Moreover, for
the actual data observation/simulation, \revI{the} integration over a range of
time/frequency will cause the loss of amplitude. This is the
well known smearing effect or decoherence. \citet{Smirnov2011}
provides
a useful first order approximation for this effect. The implementation
is relatively easy. However the corresponding time consumption
increases significantly\footnote{The total number of visibilities increases
from $N_\mathrm{IF}\cdot N_\mathrm{AP}$ to $(2N_\mathrm{IF}+1)\cdot(2N_\mathrm{AP}+1)$.}.
We plan to implement this part when the computation efficiency is
not a big issue.

Besides that, at present, the antenna response
within the FoV is assumed to be constant. This assumption is
reasonable for observations with very long baseline, since the
imaging area is significantly smaller than the FoV. However, for
wide field array, of which the imaging area is much larger, \revI{the} variation
of antenna response within the FoV must be taken into account,
so as to achieve a more realistic result.

\section{Summary}\label{sec:sum}

In this paper, we present \uvom, so as
to fulfill the requirement of VLBI simulations for both space and ground
VLBI observations. It supports various kinds of stations,
including Earth (ground) fixed, Earth orbit, Lunar fixed, Lunar orbit,
Moon-Earth and Earth-Sun Lagrange 1 and 2 points, etc. The main
functionalities of this toolkit are: (1) Trajectory calculation; (2)
Baseline $uv$ calculation; (3) Visibility simulation; (4) Image and beam
 reconstruction.

\uvom~provides two methods for visibility simulation
and image reconstruction, namely FFT and \revI{DFT}. The
latter one avoids extra artifacts introduced by the gridding process, so
as to achieve the high sensitivity necessary for neutral hydrogen studies.
Moreover, $w$ term calculation is naturally supported by this method,
which gives \uvom~the ability for simulations of wide field array.

As a demonstration of \uvom, two scenarios of completely different scales
are presented. One is space VLBI, which compares the resolutions of VLBI
networks with and without Moon-Earth baselines. Another is ground based SKA1-Mid array,
which exhibits the toolkit's capability of visibility calculation and radio
imaging for wide field arrays.

\uvom~is open for access and will be updated continuously. All the
necessary documents, examples and packages are publicly available
in GitHub repo: \url{https://github.com/liulei/omniuv}.

\begin{acknowledgments}
LL appreciates the helpful discussions with Dr. Li Guo, Dr. Zhenghao Zhu, Dr. Shanshan Zhao and Mr. Shaoguang Guo.
This work is supported by the National Natural Science Foundation of China
(Grant No. 11903067, U1938114, 11973011 and 11573057), the National SKA Program of China (Grant No. 2020SKA0110102),
the National Science and Technology Basic Conditions Platform Project
``National Basic Science Data Sharing Service Platform'' (Grant No. DKA2017-
12-02-09), the Key Technical Talents of Chinese Academy of Sciences, the Shanghai Outstanding Academic Leaders Plan, the Lunar Exploration Project and the Key Cultivation Projects of Shanghai Astronomical Observatory.

\end{acknowledgments}

\appendix
\section{Accuracy of $uvw$ calculation}
\begin{figure}
\plotone{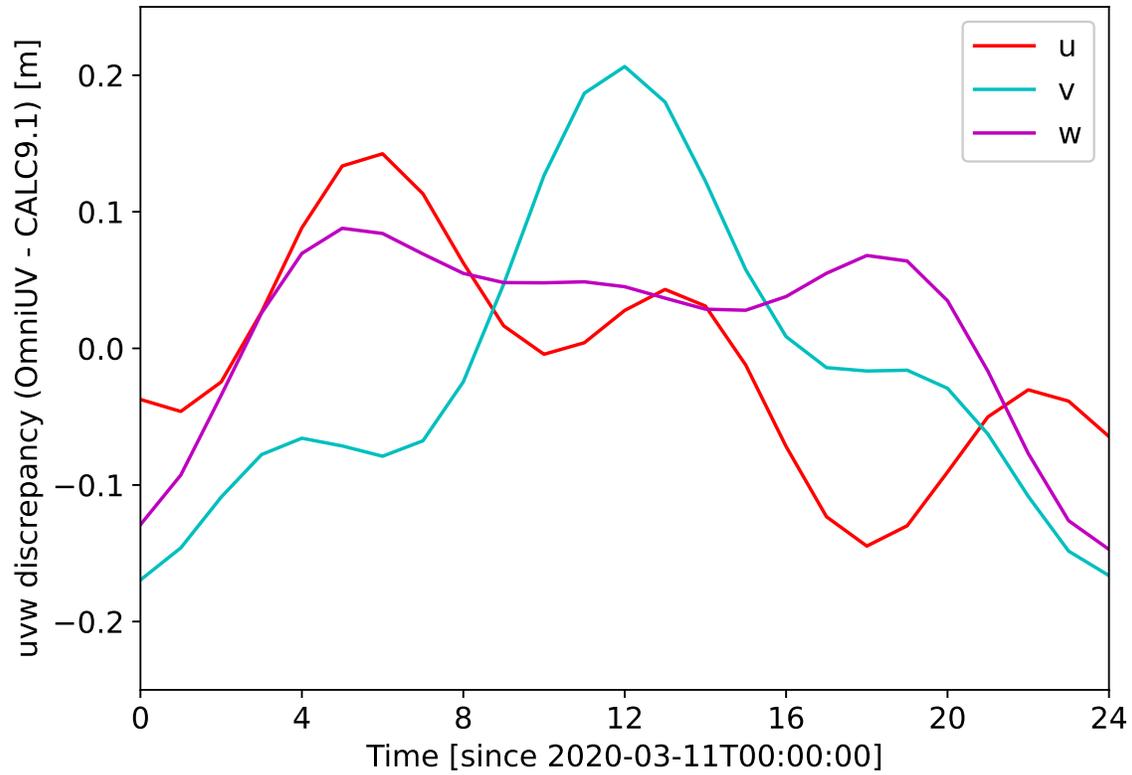}
\caption{Discrepancy of $uvw$ as a function of time. \label{fig:duvw}}
\end{figure}

To verify the accuracy of \uvom~$uvw$ calculation, we perform comparison between
\uvom~and CALC 9.1\footnote{https://space-geodesy.nasa.gov/techniques/tools/calc\_solve/calc\_solve.html}. Both calculations are set up for the SKA1-Mid site with 
identical earth orientation parameters and site coordinate. The result is
presented in Fig.~\ref{fig:duvw}.
According to the figure, the discrepancies are well within 20~cm.
We postulate the still existing discrepancy is caused by various types of 
tide effects which are not taken into account by \uvom.

\begin{figure}
\plotone{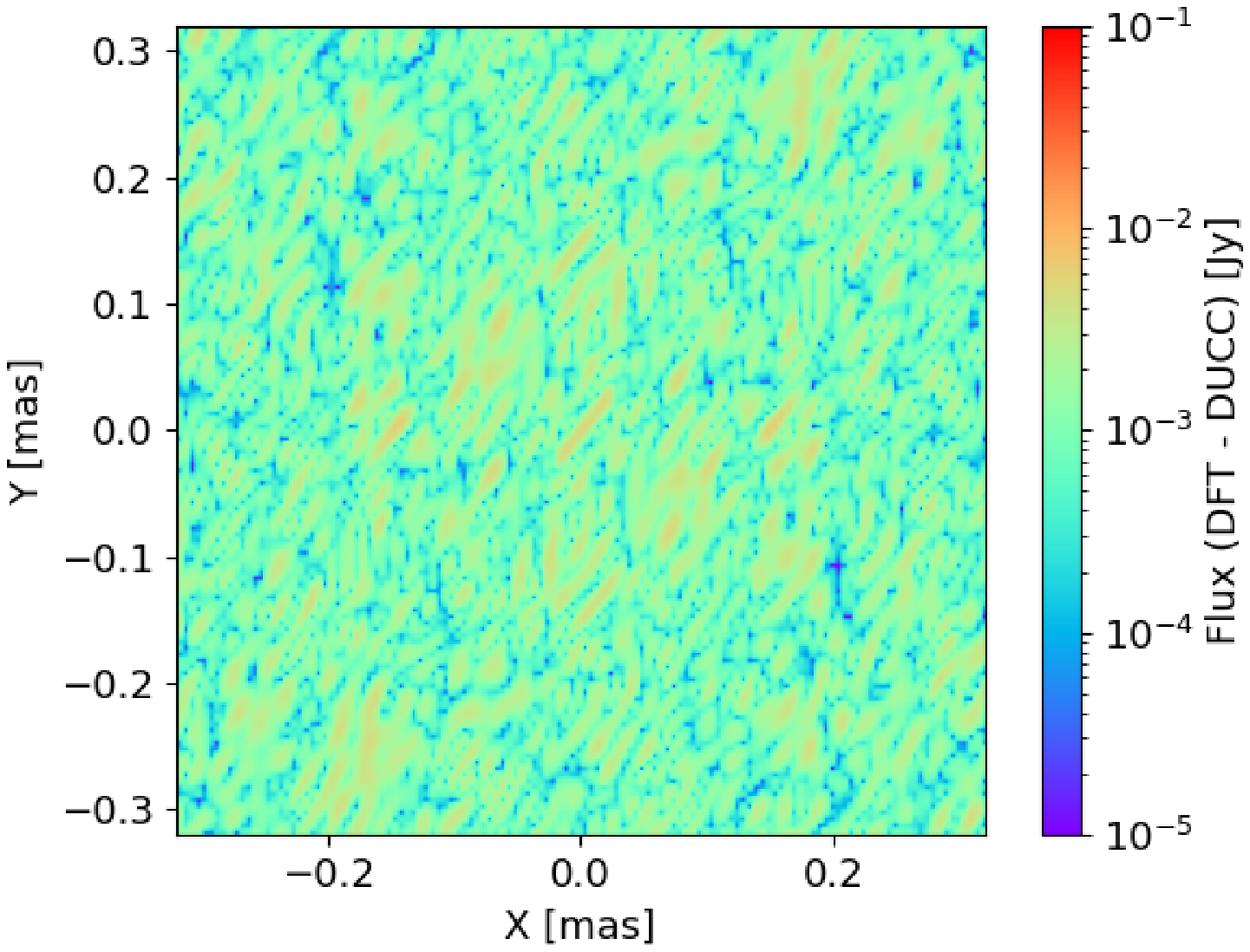}
\caption{Discrepancy of image reconstruction result between DFT and 
\texttt{DUCC} lib for the Moon-Earth network simulation in
Sec.~\ref{sec:svlbi}.\label{fig:svlbi_dft_ducc}}
\end{figure}

\begin{figure}
\plotone{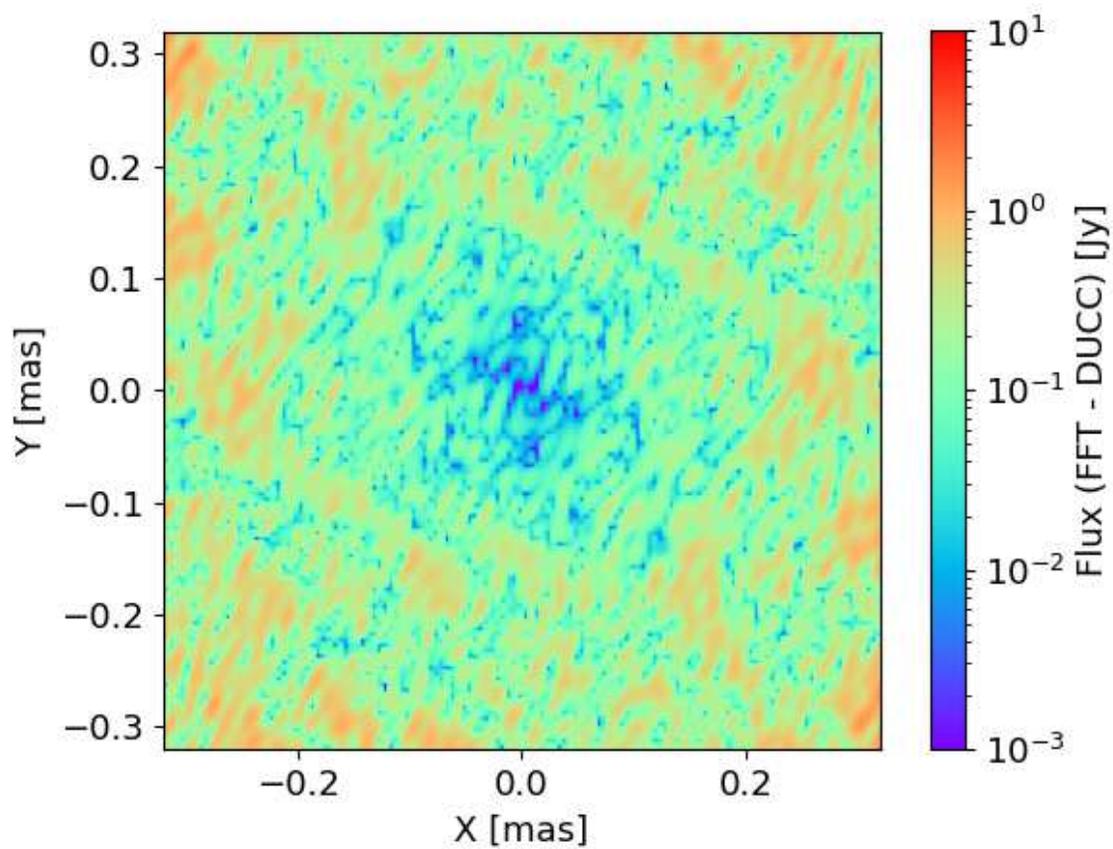}
\caption{Discrepancy of image reconstruction result between FFT (implemented
with nearest neighbour gridding in \uvom) and \texttt{DUCC} lib for the 
Moon-Earth network simulation in Sec.~\ref{sec:svlbi}.\label{fig:svlbi_fft_ducc}}
\end{figure}

\begin{figure}
\plotone{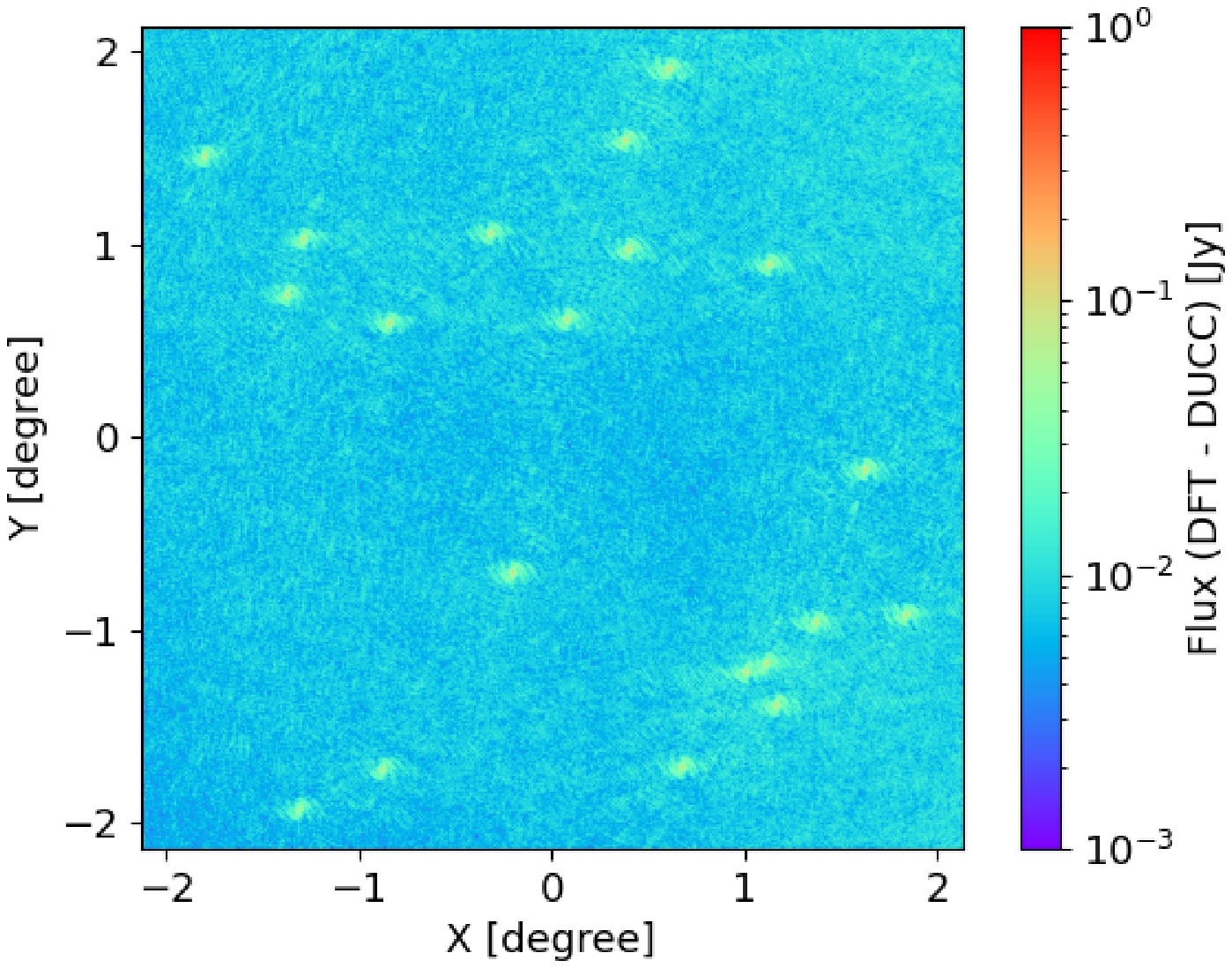}
\caption{Discrepancy of image reconstruction result between DFT 
and \texttt{DUCC} lib for the wide field SKA1-Mid simulation in
Sec.~\ref{sec:widefield}. \label{fig:ska_dft_ducc}}
\end{figure}

\section{Accuracy of image reconstruction}
We compare the image reconstruction results between \uvom~and \texttt{DUCC}
lib \citep{ducc}. The latter one implements the novel analytical kernel 
for $uv$ plane gridding, which guarantees high accuracy when transformed to 
the image plane. The comparisons for the Moon-Earth network 
are presented in Fig.~\ref{fig:svlbi_dft_ducc} and \ref{fig:svlbi_fft_ducc}. 
Here ``FFT'' refers
to the nearest neighbour gridding method in the current 
version of \uvom. Obviously its discrepancy with \texttt{DUCC} lib is much larger
than that of the DFT method. The comparison for the SKA1-Mid array is presented 
in Fig.~\ref{fig:ska_dft_ducc}. Compared with the small field imaging in
Fig.~\ref{fig:svlbi_dft_ducc}, the discrepancy is relatively larger in the 
wide field mode. FFT with nearest neighbour gridding is of high efficiency and
easy to implement. We keep it in the toolkit to facilitate code development and
testing. However, this method is not recommended in the actual application of
image reconstruction since its accuracy is low. 
In general, the accuracy of \texttt{DUCC} lib is as high as DFT, but with 
less time consumption comparable to FFT (implemented with nearest neighbour 
gridding in \uvom) for both small field and wide field imaging. We plan to
support \texttt{DUCC} lib in the future release of \uvom~and set it as the 
default scheme for image reconstruction.

\end{document}